%% file: sample-sigconf.tex
\documentclass[sigconf]{acmart}

\usepackage{booktabs} 

\setcopyright{rightsretained}

\usepackage{algorithm, algorithmicx} 
\usepackage{algpseudocode} 

\usepackage{amssymb}
\usepackage{amsmath}

\usepackage{multirow}
\usepackage{graphicx}
\usepackage{subfigure}

\acmDOI{10.475/123_4}

\acmISBN{123-4567-24-567/08/06}

\acmConference[WOODSTOCK'97]{ACM Woodstock conference}{July 1997}{El
  Paso, Texas USA} 
\acmYear{1997}
\copyrightyear{2016}

\acmArticle{4}
\acmPrice{15.00}

\begin{document}
\title{Beyond Precision: A Study on Recall of Initial Retrieval with Neural Representations}

\author{Yan Xiao, Jiafeng Guo, Yixing Fan, Yanyan Lan, Jun Xu, and Xueqi Cheng}
\affiliation{
  \institution{University of Chinese Academy of Sciences, Beijing, China\\
  CAS Key Lab of Network Data Science and Technology, Institute of Computing Technology,\\ Chinese Academy of Sciences, Beijing, China
  } 
}
\email{xiaoyanict@foxmail.com, fanyixing@software.ict.ac.cn, {guojiafeng, lanyanyan, junxu, cxq}@ict.ac.cn}

\renewcommand{\shortauthors}{X. Yan et al.}

\begin{abstract}
Vocabulary mismatch is a central problem in information retrieval (IR), i.e., the relevant documents may not contain the same (symbolic) terms of the query. Recently, neural representations have shown great success in capturing semantic relatedness, leading to new possibilities to alleviate the vocabulary mismatch problem in IR. However, most existing efforts in this direction have been devoted to the re-ranking stage. That is to leverage neural representations to help re-rank a set of candidate documents, which are typically obtained from an initial retrieval stage based on some symbolic index and search scheme (e.g., BM25 over the inverted index). This naturally raises a question: if the relevant documents have not been found in the initial retrieval stage due to vocabulary mismatch, there would be no chance to re-rank them to the top positions later. Therefore, in this paper, we study the problem how to employ neural representations to improve the recall of relevant documents in the initial retrieval stage. Specifically, to meet the efficiency requirement of the initial stage, we introduce a neural index for the neural representations of documents, and propose two hybrid search schemes based on both neural and symbolic indices, namely the parallel search scheme and the sequential search scheme. Our experiments show that both hybrid index and search schemes can improve the recall of the initial retrieval stage with small overhead.

\end{abstract}

%
%



\maketitle

\input{samplebody-conf}

\bibliographystyle{ACM-Reference-Format}
\bibliography{ref} 

\end{document}

%% file: samplebody-conf.tex
\section{Introduction}
IR pipeline in modern search systems typically consists of two stages \cite{dang2013two}, namely the initial retrieval stage and the re-ranking stage. The initial retrieval stage aims to retrieve a small subset from the whole corpus that contains as many relevant documents as possible (i.e., high recall) with small cost (i.e., high efficiency). Without loss of generality, this is usually achieved under a symbolic index and search scheme in modern search systems. For example, a term-based inverted index is first built over the corpus, and some simple term-based ranking algorithm, like BM25\cite{robertson1994some}, is employed to efficiently find the candidate subset. The re-ranking stage, on the other hand, aims to produce a high-quality (i.e., high precision) ranking list of the subset. Since the subset usually contains much fewer documents than the whole corpus, more complicated ranking algorithms, such as learning to rank algorithms \cite{freund2003efficient, joachims2002optimizing} or deep neural models \cite{guo2016deep, pang2017deeprank} , could be involved in this stage for the re-ranking task.

The above pipeline has been widely adopted in most practical search systems, and a long-standing challenge it faces is the vocabulary mismatch problem, i.e., the relevant documents may not contain the same (symbolic) terms of the query.  While there have been many efforts in developing ranking algorithms to address this challenge \cite{xu2011kernel, guo2016semantic, mitra2016dual}, most of them were at the re-ranking stage. But what if the relevant documents have not been found in the initial retrieval stage due to vocabulary mismatch, which is very likely to happen due to the symbolic index and search scheme? In this case, there would be no chance to re-rank those missing relevant documents to the top positions later. Therefore, we argue that it is critical to tackle the challenge at the very beginning. In other words, we shall attempt to solve the vocabulary mismatch problem at the initial retrieval stage, rather than only addressing it at the re-ranking stage.

In recent literature, neural representation (e.g., word embedding \cite{mikolov2013distributed}) has achieved great success in capturing the semantic relatedness. By representing each word as a dense vector, similar words would be close to each other in the semantic space and the linguistic relations between words could be simply calculated via algebra. Such neural representations bring new possibility to alleviate the vocabulary mismatch problem in IR beyond the traditional symbolic term-based representation. Unfortunately, most existing efforts \cite{guo2016semantic, mitra2016dual, guo2016deep} in using neural representations for IR have been devoted to the re-ranking stage .

In this paper, we study the problem how to employ neural representations to improve the recall of relevant documents in the initial retrieval stage. To address this problem, we need to solve two major challenges, i.e., how to index neural representation of documents and how to search with neural index. To meet the efficiency requirement of the initial stage, we represent each document as a weighted sum of word embeddings, and introduce a $k$-nearest-neighbor ($k$-NN) graph based neural index which is efficient in both construction and search over dense vectors. We then propose two hybrid search schemes based on both neural and symbolic indices, namely the parallel search scheme and the sequential search scheme. The parallel search scheme retrieves documents based on symbolic index and neural index simultaneously, and merges the top results together to obtain the candidate subset. In this way, both the symbolic and neural indices act as two separate memories of the corpus. In the sequential search scheme, we first retrieve seed documents based on the symbolic index, and then expand the candidate subset based on the neural index. In this case, the symbolic index acts as the precise memory while the neural index acts as the associative memory.

To evaluate the effectiveness and efficiency of our proposed hybrid index and search schemes, we conduct extensive experiments on two IR benchmark collections. The experiments show that by using neural index and hybrid search scheme, we could improve recall with small overhead for initial retrieval. Among the two search schemes, the sequential search scheme can achieve better recall than the parallel search scheme at some additional cost. We further conduct detailed analysis of the retrieval results and find that symbolic index and search scheme play an important role while the neural scheme could provide complementary results of relevant document as compared to the symbolic scheme.

Overall, the major contributions of our work are as follows:
\begin{itemize}
\item We propose to enhance recall at the initial retrieval stage with neural representations. We introduce a $k$-NN graph based neural index and further propose two hybrid search schemes, i.e., the parallel search scheme and the sequential search scheme.
\item We conduct extensive experiments on two IR benchmark collections to evaluate the effectiveness and efficiency of our proposed approaches for initial retrieval.
\item  We conduct detailed analysis to study the utility and difference of both symbolic and neural indices.
\end{itemize}

The rest of the paper is organized as follows. Section 2 gives a brief summary of related work. We describe the detailed implementation of hybrid index and search schemes in Section 3. In Section 4 we present the experimental results and conduct detailed analysis. Section 5 concludes the paper and talks about the future work.
 
\section{Related work}
In this section, we introduce the related work, including the existing methods for initial retrieval and explorations of neural representations for IR.

\subsection{Initial Retrieval}
Conventional initial retrieval relies on an inverted index to obtain a list of document candidates, and then simple model such as BM25 can be fast executed over these candidates to retrieve initial results. Two-stage learning to rank \cite{dang2013two} was proposed to replace BM25 with a more complex ranker. This ranker is learned beyond query terms, including weighted phrases, proximities, and expansion terms. However, this has not alleviated the vocabulary mismatch problem since candidates are still obtained from the inverted index, and the efficiency has not been taken into consideration in this work either.

Other efforts have attempted to replace the symbolic inverted index with $k$-NN search. \citet{li2014two} proposed the two-stage hashing scheme for fast document retrieval. In their work, they represent both query and document as TF-IDF weight vectors, and use the cosine similarity to evaluate the similarity between query and document. They perform $k$-NN search using Locality Sensitive Hashing (LSH) \cite{datar2004locality} to retrieve document candidates in the first stage and then re-rank these documents using Iterative Quantization \cite{gong2013iterative}. The two-stage hashing scheme can be more efficient than traditional IR baselines, but has not achieved the same effectiveness. Meanwhile, they evaluated by the precision while not by the recall.

To address the vocabulary mismatch problem, \citet{boytsov2016off} proposed to use complex initial retrieval model and perform $k$-NN search in non-metric space. They use the linear combination of BM25 and IBM Model I \cite{brown1993mathematics} as the non-metric similarity, and find pivot-based index Neighborhood Approximation (NAPP) \cite{tellez2013succinct} can achieve some good results on question answering (QA) datasets. NAPP selects several documents as the pivots, and each query and document is represented by its $k$ nearest pivots ($k$-NPs) computed by the brute force search. Given a query, the documents sharing a pre-specified number of $k$-NPs with the query are filtered to compute real distance. However, $k$-NN search in non-metric space leads to new indexing challenges, i.e., many hand-crafted optimizations and heuristic computations are needed for search efficiency\cite{boytsov2016off}. On the other hand, the effectiveness and efficiency of non-metric $k$-NN search have not been evaluated on the initial retrieval for IR, which is different with QA. Note that they have performed $k$-NN search using cosine similarity between averaged word embeddings of questions and answers, but this can not achieve good effectiveness.

\subsection{Neural Representations for IR}
Existing work of neural representations for IR mainly explore two ways, i.e., leveraging word embeddings to enhance the representations of query and document\cite{guo2016deep, mitra2016dual}, or learning query and document representations by a deep model\cite{huang2013learning}.

\citet{guo2016deep} proposed to build local interactions between each pair of words from a query and a document based on word embeddings. In their work, the local interactions are mapped into a fixed-length matching histogram, and then this histogram serves as the input of a deep model to learn the relevance matching score.\citet{mitra2016dual} proposed a new ranking method based on comparing each query word to a centroid of the document word embeddings. In their work, the word embeddings used for query and document are in dual embedding space. Their model is effective in re-ranking top documents returned by a search engine, and a linear mixture of their model and BM25 can be employed to rank a larger set of candidate document. 

Rather than leveraging word embeddings directly, \citet{huang2013learning} proposed to learn query and document representations from clickthrough logs by a deep model and model the relevance by cosine similarity between query and document representations.

All these work have shown that employing neural representations for IR can achieve better effectiveness. However, these work aim at the re-ranking stage, while not the initial retrieval stage. If a relevant document is not contained in the initial results, the re-ranking stage can only generate a sub-optimal ranking list. 

\section{Our Approach}
In this section, we firstly introduce the symbolic index and neural index used in our approach, and then describe the parallel and sequential search schemes based on these two kinds of indices. Finally, we give some discussions on these two search schemes.

\subsection{Symbolic Index}

\begin{figure}[h]
    \centering
    \includegraphics[scale=0.4]{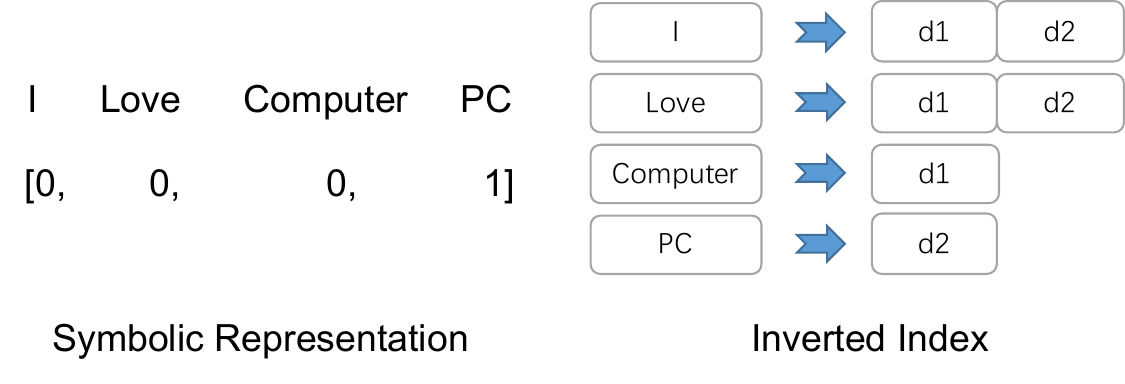}
    \caption{Symbolic representation and index.}
    \label{fig:symbolic}
\end{figure}

Conventional IR is based on symbolic term-based representation, where query and document are represented as a bag of terms (i.e., Bag-of-Words representation), and each term is represented as a one-hot vector. Each dimension of such a symbolic representation denotes the occurrence times of one distinct term, and is treated as independent from others. Due to the sparsity of symbolic representation, inverted index is widely employed as the core index structure in model search systems. In an inverted index, each term is linked to a posting list of its occurrence information in the corpus, including the document identifier, corresponding frequency and so on. For example, the document only containing $PC$ is represented as [0,0,0,1] as shown in Figure~\ref{fig:symbolic}.

Given a corpus, the documents are scanned one by one to obtain their symbolic representations and then it is efficient to construct an inverted index. Specifically, for each non-zero dimension of a symbolic representation, the occurrence information is appended to the posting list of the corresponding term.

\subsection{Neural Index}

In this work, we further employ neural representation for documents so that we can search documents beyond traditional symbolic terms to enhance recall at the initial retrieval stage. Rather than learning a representation for each document, we simply adopted the TF-IDF weighted sum of word embeddings as its neural representation due to the trade-off between effectiveness and computation efficiency\cite{mitra2016dual, boytsov2016off}. Since the neural representations are continuous and dense vectors, the inverted index is not suitable for searching over them.
 
\begin{figure}[ht]
    \centering
    \includegraphics[scale=0.4]{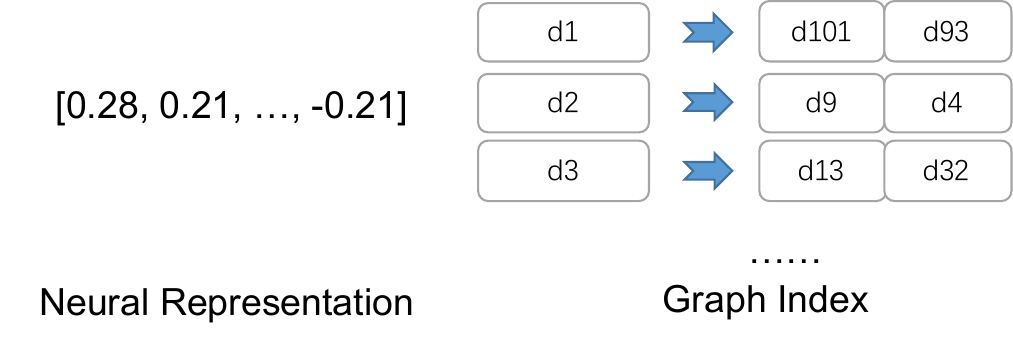}
    \caption{Neural representation and index.}
    \label{fig:neural}
\end{figure}

Indexing technologies for dense vectors have been widely studied and many structures have been proposed, such as $k$-$d$ tree\cite{bentley1975multidimensional} , LSH \cite{datar2004locality}, neighborhood graph-based methods \cite{hajebi2011fast, malkov2016efficient}. 

Here we adopt a $k$-nearest neighbor ($k$-NN) graph \cite{hajebi2011fast} based structure as the neural index, which is shown in Figure~\ref{fig:neural}. Each document is linked to its $k$ most semantically similar documents measured by cosine similarity between their neural representations. The links can be reversed to obtain undirected $k$-NN graph\cite{li2016approximate}, i.e., attaching the reversed neighbors behind the $k$-NNs for each document.

Since constructing a precise $k$-NN graph is time-consuming for a large scale dataset, many approximate algorithms\cite{dong2011efficient, chen2009fast} have been proposed. We adopt the state-of-the-art algorithm NN-Descent \cite{dong2011efficient}, which can be extremely efficient to construct a highly precise $k$-NN graph. In an evolving dataset, the neighbors will change with insertions. Rather than rebuilding the whole graph, we could perform approximate nearest neighbor search to find the k-NNs for a new inserted point and update the neighbors of these k-NNs with the new inserted point. 


\subsection{Parallel Search Scheme}
With both the symbolic and neural indices in hand, a natural idea is to retrieve based on these two kinds of indices simultaneously, and merge the top results together to obtain the candidate subset. In this way, both the symbolic and neural indices act as two separate memories of the corpus. 

\begin{figure}[h]
    \centering
    \includegraphics[scale=0.35]{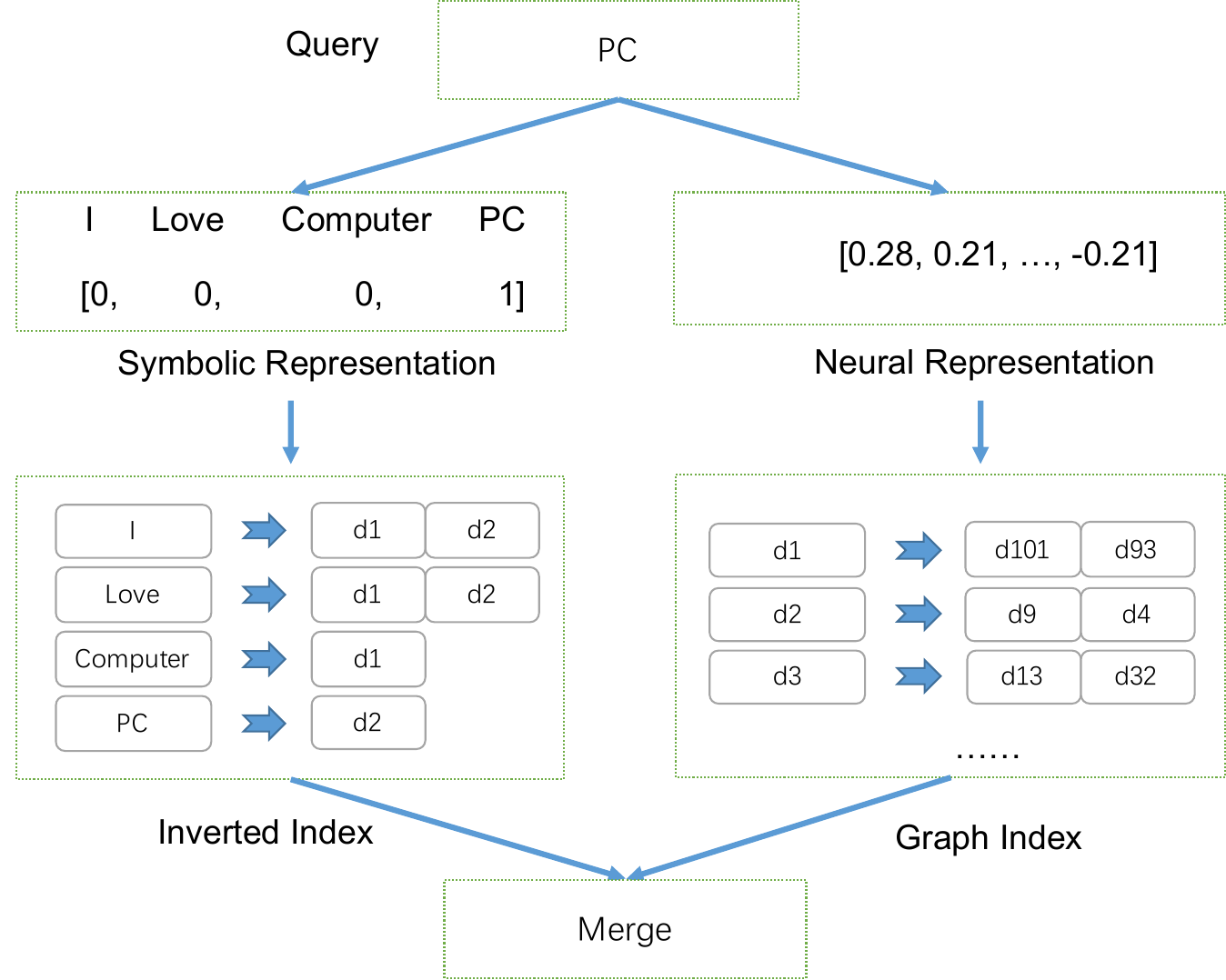}
    \caption{Parallel search scheme.}
    \label{fig:parallel}
\end{figure}

The search process is shown in Figure~\ref{fig:parallel}. As there are two search paths, we name this scheme as the parallel search scheme (ParSearch in short).

The symbolic search is conducted in the left path, where posting lists of the corresponding query terms are first looked up through the inverted index and merged to obtain the candidate documents. Then we use BM25 to retrieve $m$ candidates from these documents, where $m$ is pre-specified. Simultaneously, in the right path, the neural search is performed to find $n$ nearest neighbors of the query in the semantic space via traversing on the graph index\cite{hajebi2011fast}, where $n$ is the number of candidate documents for initial retrieval. Specifically, some documents are sampled as starting points, and the linked documents of the most similar and unexplored document are iteratively explored to approach the query. At each iteration, $n$ most similar documents to the query are kept. The similarity between query and document is measured by cosine similarity between their neural representations. 

The results of symbolic search (i.e., $m$ candidates by BM25) are based on exact matching signals, while the results of neural search (i.e., $n$ candidates by cosine similarity) are based on semantic matching signals. Since these two kinds of documents are not scored in the same space and symbolic search is more precise than neural search as shown in previous work \cite{boytsov2016off, mitra2016dual}, thus we introduce a new aggregation method to get final results. Specifically, we take the set of documents from the symbolic search as the base, and merge those from the neural search into it to obtain sufficient number of candidate documents. In the merging process, we will scan the documents from neural search from top to end one by one, and if it is not in the symbolic candidate set, it will be merged. This merging process is efficient since we do not need to compute any score or re-rank any document. This is reasonable as we only care about the recall rather than the precision. 

The parallel search scheme is formally described in Algorithm~\ref{alg:parallel_search}.

\begin{algorithm}
\scriptsize
    \caption{Parallel Search Scheme}  
    \label{alg:parallel_search}
    \begin{algorithmic}\small
    	\Require query q, number of initial candidate documents $n$, number of candidate documents retrieved by symbolic search $m$
        \Ensure $n$ initial candidate documents
        
        \State $M \leftarrow m$ documents based on the inverted index and BM25
        \State $C \leftarrow n$ documents based on the undirected $k$-NN graph index and cosine similarity
          
        \For{\textbf{each} $d \in C$}
        	\If {$d \in M$ }
        		\State continue
        	\EndIf
        	
        	\State $M \leftarrow M \cup \{d\}$  
        	\If {$|M| == n$}
        		\State break
        	\EndIf  	
        \EndFor
        \State return $M$
    \end{algorithmic}
\end{algorithm}

\subsection{Sequential Search Scheme}
Although parallel search scheme is straightforward and efficient, it may not be very effective since the retrieval performance of neural representation alone is questionable. Another possible way to combine symbolic and neural search is a sequential manner, with the expectation that documents semantically similar to relevant documents are expected to be relevant. In this way, we can first employ symbolic search to find some candidate documents as the seeds, and then use neural search to expand the seeds to obtain the final candidate subset. In this case, the symbolic index acts as the precise memory while the neural index acts as the associative memory of the corpus. 

\begin{figure}[ht]
    \centering
    \includegraphics[scale=0.3]{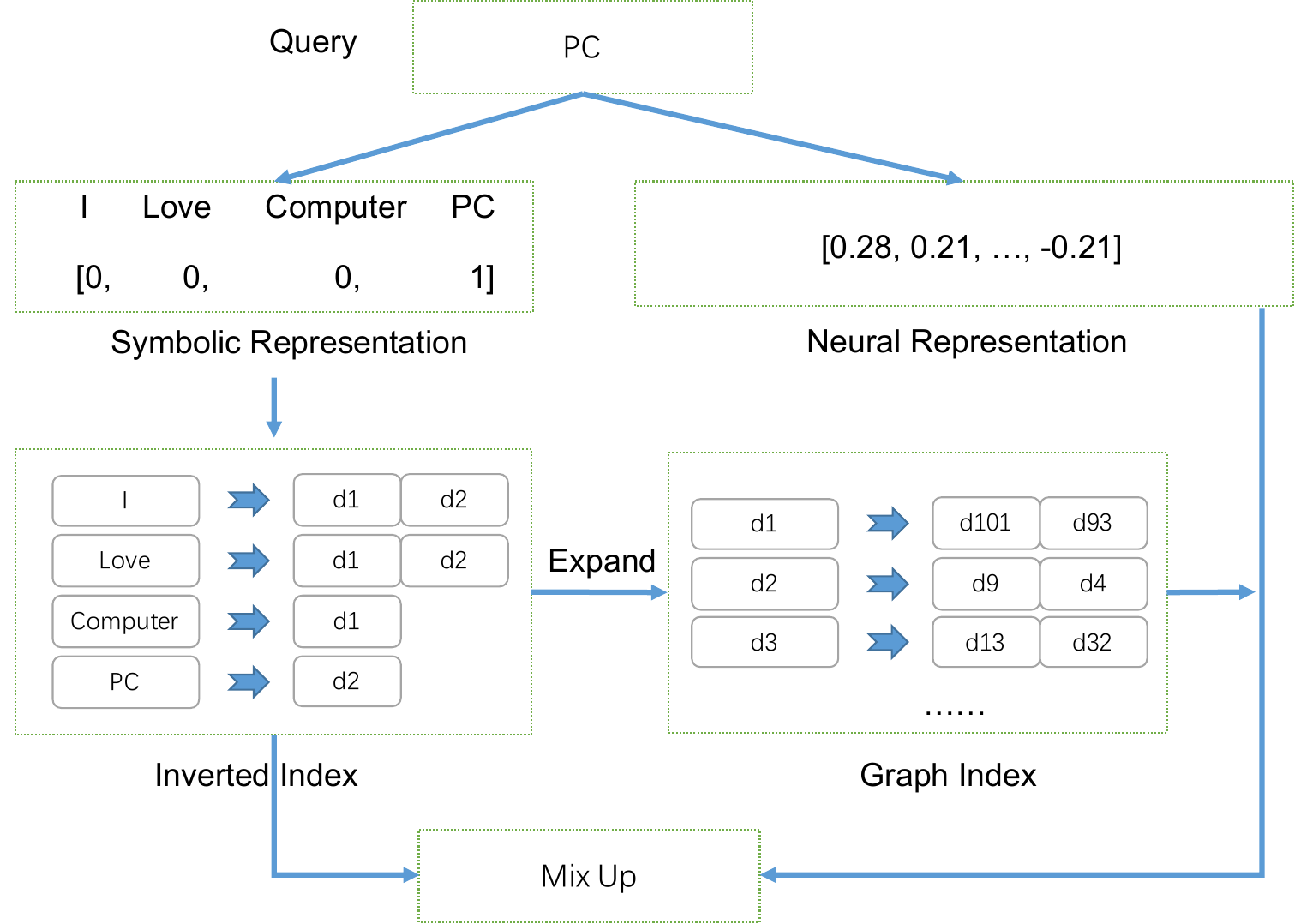}
    \caption{Sequential search scheme.}
    \label{fig:sequential}
\end{figure}

The search process is shown in Figure~\ref{fig:sequential}. As there is only one chained search path, we name this scheme as the sequential search scheme (SeqSearch in short).

Given a query, we firstly conduct symbolic search to retrieve seed documents as the same with ParSearch. We expand these seeds to associate their semantically similar documents based on the graph index, i.e., obtaining their $k$-NNs via looking up the graph. Then cosine similarity is used to obtain the top associated documents. We return the seeds with the later retrieved candidates together as the final results. 

To control the efficiency overhead from neural search, we further propose a heuristic strategy to reduce the computation, i.e., only expanding a pre-specified proportion of highly scored seed documents. The underlying intuition is that $k$-NNs of a seed are more relevant if the seed has higher exact matching score. We do not reduce the number of seeds since exact matching is the most important signal in IR\cite{guo2016deep} and more reliable than semantic matching.

The sequential search scheme is formally described in Algorithm~\ref{alg:sequential_search}. 

\begin{algorithm}
\scriptsize
    \caption{Sequential Search Scheme}  
    \label{alg:sequential_search}
    \begin{algorithmic}\small
    	\Require query q, $k$-NN graph $G$, number of initial candidate documents $n$, number of seed documents $s$, expanding proportion $p$
        \Ensure $n$ initial candidate documents
        
        \State $S \leftarrow$ $s$ seed documents based on the inverted index and BM25
        \State $D \leftarrow \emptyset$
        \For{\textbf{each} $d \in$ $S[1:p*s]$}
        	\State $U \leftarrow$ $k$ neighbors of $d$ in $G$
        	\State $D \leftarrow D \cup \{U\setminus S\}$       	
        \EndFor
        \State $N \leftarrow$ Retrieve $n-s$ documents from $D$ based on cosine similarity
        \State return $S \cup  N$
    \end{algorithmic}
\end{algorithm}

\subsection{Discussions}
The ParSearch and SeqSearch both employ neural representations to improve the recall of relevant documents in the initial retrieval stage. Here we make detailed discussions on them.

\subsubsection*{Efficiency}
The ParSearch considers symbolic search and neural search separately, which can be efficient since these two processes can be conducted at the same time. In our experiments, we find that the neural search executes more efficiently than symbolic search. Thus the ParSearch only has small and almost constant overhead from the merging process.

As for the SeqSearch, neural search is executed after symbolic search, thus may sacrifice some efficiency. Note here in SeqSearch, we only need to look up $k$-NNs for each seed document, where the computational complexity is O(1). Therefore, the efficiency overhead of neural search is almost constant and has no relationship with the query length and collection size. With the increasing of query length or corpus size, an increasing number of documents are likely to contain the terms of the query, leading to more cost of symbolic search (i.e., more documents are initially retrieved by inverted index and scored by BM25). The relative overhead from neural search would become negligible as compared with the symbolic search.

\subsubsection*{Effectiveness}
Although it is more efficient to retrieve separately as ParSearch does, this may have a disadvantage against the effectiveness since semantic matching with neural representation alone may bring many noise matching signals\cite{pang2017deep}, e.g., any two adjective terms could contribute some matching signals. 

Comparing with the ParSearch, neural search of SeqSearch is based on symbolic search, i.e., the semantically matched document are retrieved from the documents associated by exactly matched seeds. The association process could filter plenty of noise documents and ensure the quality of neural search. Our experiments have also demonstrated this.

\subsubsection*{Extendibility}
The $k$-NN graph is organized by linking each document to its $k$ most similar documents. We can define the similarity as a more powerful function, such as the linear combination of a exact matching model and a semantic matching model. Such a similarity function can be neither metric nor symmetric, but could be supported by NN-Descent\cite{dong2011efficient}. 

Since performing $k$-NN search with non-metric graph is more complex and challenging \cite{boytsov2016off} in ParSearch than  association in SeqSearch, it seems SeqSearch has better extendibility than ParSearch. The major reason lies in that the ParSearch relies on the neural index to find $k$-NNs for unseen queries, while the SeqSearch only need a neural index to look up $k$-NNs for existing documents. 

\section{Experiments}
In this section, we conduct experiments to evaluate the effectiveness and efficiency of ParSearch and SeqSearch on two IR benchmark collections.

\subsection{Datasets}
\begin{table}[h]
\centering
\caption{Statistics of the TREC collections used in
this study.}
\label{tab:datasets}
\begin{tabular}{ccc}
\toprule
 & Robust04 & WT2G \\
\midrule
Vocabulary & 0.6M & 0.97M \\
Document Count & 0.5M & 0.25M \\
Collection Length  & 252M & 261M \\
Query Count & 250 & 50 \\
\bottomrule
\end{tabular}
\end{table}

We conduct experiments on two TREC collections, Robust04 and WT2G. The details of the collections are provided in Table~\ref{tab:datasets}. Robust04 is a news dataset and its topics are collected from TREC Robust Track 2004. WT2G is a general Web crawl and its topics are collected from TREC Web Track 1999. We make use of both the title and the description of each TREC topic in our experiments. Both documents and queries are white-space tokenized, lowercased, and stemmed using the Krovetz stemmer \cite{krovetz1993viewing}. Stop word removal is performed on query words during retrieval using the INQUERY stop list \cite{callan1995trec}.

\subsection{Baselines and Experimental Settings}
We adopt the symbolic index based method and neural index based method as the baselines. We also conduct the experiment on non-metric $k$-NN search\cite{boytsov2016off} using the linear combination of BM25 and cosine similarity as the matching score. These methods include:

\textbf{BM25}: The BM25 formula \cite{robertson1994some} is a highly effective symbol based initial retrieval model. We use the inverted index implemented in Apache Lucene \footnote{http://lucene.apache.org}. We tune the parameters $k1$ and $b$ to obtain the bast performance. There still some other approaches relying on BM25, such as our approaches, and we use the same parameter settings as the BM25 baseline.

\textbf{Cosine}: The matching score is the cosine similarity between neural representations of query and document \cite{brokos2016using, mitra2016dual}. For fast initial retrieval, approximate $k$-NN search is performed as in \cite{boytsov2016off} . We adopt undirected 20-NN graph \cite{hajebi2011fast, li2016approximate} as the index, whose basic idea is to iteratively explore the neighbors of currently kept nearest candidates to approach the query. Since the performance of this method depends on the trade-off between efficiency and effectiveness, we tune the number of kept nearest candidates to control the trade-off. We use subscript to denote this number.

\textbf{LinComb}: The matching score is the linear combination of BM25 score and cosine similarity. Given the query $q$ and a document $d$, the matching score is defined as below,

\begin{equation}
	\begin{aligned}
    Matching\ Score(q, d) &= \lambda*BM25(q, d) \\
       	&+ (1-\lambda)*Cosine(\vec{q}, \vec{d})  \notag
    \end{aligned} 
\end{equation}
where $\lambda$ denotes the co-efficiency balancing the two scores. NAPP \cite{tellez2013succinct} is adopted as the index for non-metric $k$-NN search as used in \cite{boytsov2016off}. We implement the LinComb method based on the public code\footnote{https://github.com/nmslib/nmslib/tree/nmslib4qa\_cikm2016} and adopt the same optimization technologies used in \cite{boytsov2016off}, such as using pseudo-documents containing 1,000 terms sampled from the set of 50,000 most frequent terms as the pivots. For each document, we pre-compute and store the BM25 score of each term in it for fast online search. To choose the $\lambda$, we perform a parameter sweep between 0.00 and 1.00 at intervals of 0.01 using brute force. We tune the number of shared $k$-NPs between query and documents to control the effectiveness and efficiency trade-off. We use subscript to denote the number of shared $k$-NPs.

\textbf{ParSearch}: For our proposed parallel search scheme, we adopt the same index as Cosine for neural representations.

\textbf{SeqSearch}: As for our proposed sequential search scheme, we use the same 20-NN graph in Cosine. We use subscript to denote the expanding proportion $p$.

The word embeddings used for neural representations are trained on Robust04 and WT2G collections respectively by the Skip-Gram model implemented in Word2Vec \footnote{https://code.google.com/p/word2vec/}. Specifically, we set $min$-$count=0$ to keep all the words and use 200-dimension embeddings.

The experiments on Robust04 and WT2G are conducted on a machine with 2.7GHz Intel Core i5-5257U CPU and 8GB memory, and a single thread is used to test initial retrieval performance for all the methods. The Cosine and LinComb methods are implemented in C++11 and and compiled with Clang using O3 optimization flag. Their cosine similarity is computed with well-known Eigen3 \footnote{http://eigen.tuxfamily.org/index.php?title=Main\_Page} to speed up the vector dot product computation. Our proposed SeqSearch is implemented based on Apache Lucene in Java, and we simply compute the vector dot product using naive brute force since we have not found efficient library like Eigen3 for Java. This implementation is a little unfair to our SeqSearch.                   

The neural indices of Cosine, ParSearchWe and SeqSearch are based on the same 20-NN graph. Constructing the graph costs 644s and 334s on Robust04 and WT2G respectively using 4 threads.

\begin{table*}[h]
\centering
\caption{Comparison of different index and search schemes. Significant improvement or degradation with respect to BM25 for recall@1000 is indicated ($+$/$-$)($p$-$value \leq 0.05$).}
\label{tab:text_recall}
\begin{tabular}{c c c l l l l l l}
\multicolumn{9}{c}{Robust04 collection} \\
\hline \hline
& & & \multicolumn{3}{c}{Topic titles} & \multicolumn{3}{c}{Topic descriptions} \\\cline{4-6}\cline{7-9}
Index Type &  Index Method & Search Method & Recall(\%) & Ratio(\%) & Time(ms) & Recall(\%) & Ratio(\%) & Time(ms)\\
\hline
Symbolic& Inverted Index & BM25 & 68.36 & 57.84 & 4.50 & 66.68 & 53.21& 10.34 \\

\hline
\multirow{4}{2cm}{\centering Neural} & \multirow{4}{3.5cm}{\centering   Graph Index} 
& Cosine$_{100}$ & 47.76$^-$ & 40.80 & 1.01  & 49.70$^-$ & 39.83& 1.17 \\
& & Cosine$_{500}$ & 49.89$^-$ & 42.64 & 2.64  & 50.14$^-$ & 39.61 & 2.99 \\
& & Cosine$_{800}$ & 50.27$^-$ & 42.67 & 3.54  & 50.14$^-$ & 39.47 & 4.04 \\
& & Cosine$_{1000}$ & 50.37$^-$ & 42.80 & 4.17 & 50.31$^-$ & 39.57& 4.73 \\

\hline
\multirow{7}{2cm}{\centering Neural and Symbolic} & \multirow{3}{3.5cm}{\centering NAPP} 
& LinComb$_5$ & 62.13$^-$ & 53.98& 163.59 & 62.51$^-$ & 51.48& 212.40 \\
& & LinComb$_{10}$ & 28.15$^-$ & 27.27& 59.15 & 30.35$^-$ & 27.89& 88.61 \\
& & LinComb$_{15}$ & 16.23$^-$ & 16.21& 28.15 & 17.01$^-$ & 16.30& 42.35 \\

\cline{2-9}
& \multirow{4}{3cm}{\centering Inverted Index and Graph Index}
& ParSearch & 72.06$^+$ & 61.45& 5.14  &  68.46$^+$ & 55.28& 10.58 \\ 
& & SeqSearch$_{25\%}$ & 72.34$^+$ & 62.05& 5.45  & 69.15$^+$ & 55.92& 11.86 \\
& & SeqSearch$_{50\%}$ & 72.02$^+$ & 61.80& 6.37  & 69.02$^+$ & 55.87& 13.06 \\
& & SeqSearch$_{100\%}$ & 71.87$^+$ & 61.76& 8.30  & 68.83$^+$ & 55.84& 15.20 \\

\hline

\\

\multicolumn{9}{c}{WT2G collection} \\
\hline \hline
& & & \multicolumn{3}{c}{Topic titles} & \multicolumn{3}{c}{Topic descriptions} \\\cline{4-6}\cline{7-9}
Index Type & Index Method & Search Method & Recall(\%) & Ratio(\%) & Time(ms) & Recall(\%) & Ratio(\%) & Time(ms)\\
\hline
Symbolic & Inverted Index & BM25 & 82.03 &82.23 & 6.38 & 78.81 & 77.62& 11.90 \\

\hline
\multirow{4}{2cm}{\centering Neural} & \multirow{4}{3.5cm}{\centering   Graph Index} 
& Cosine$_{100}$ & 44.19$^-$ & 42.21& 0.84   & 45.76$^-$ &43.62& 0.96 \\
& & Cosine$_{500}$ & 50.04$^-$ & 52.17&2.20   & 51.11$^-$ & 49.98& 2.36 \\
& & Cosine$_{800}$ & 50.65$^-$ & 52.70&2.98   & 51.29$^-$ & 50.11& 3.16 \\
& & Cosine$_{1000}$ & 50.45$^-$ & 52.52&3.84  & 51.89$^-$ & 50.55& 3.54 \\

\hline
\multirow{7}{2cm}{\centering Neural and Symbolic} & \multirow{3}{3.5cm}{\centering NAPP} 
& LinComb$_5$ & 69.07$^-$ & 70.34& 116.43 &  71.04$^-$ & 71.92& 138.06 \\
& & LinComb$_{10}$ & 18.13$^-$ & 19.31& 44.09  & 20.59$^-$ & 23.08& 56.69 \\
& & LinComb$_{15}$ & 3.83$^-$ & 5.05& 16.23 &  4.98$^-$ & 6.41& 20.90 \\

\cline{2-9}
& \multirow{4}{3cm}{\centering Inverted Index and Graph Index} 
& ParSearch & 83.70 & 84.07& 7.84  & 79.95$^+$ & 79.03& 12.30 \\
& & SeqSearch$_{25\%}$ & 83.56 &84.12& 8.30  & 80.11$^+$ & 79.38& 13.32 \\
& & SeqSearch$_{50\%}$ & 83.58 & 84.07& 9.14  & 80.12$^+$ & 79.42& 14.42 \\
& & SeqSearch$_{100\%}$ & 83.88 & 84.29& 11.02  & 80.20$^+$ & 79.51& 16.10 \\

\hline
\end{tabular}
\end{table*}

\subsection{Evaluation Methodology}
To evaluation the effectiveness and efficiency of different index and search schemes for initial retrieval, we use averaged recall@1000 and time as the metrics. Since queries have different numbers of relevant document, we also report the ratio of all retrieved relevant documents. The averaged recall and ratio are described as follows.

\begin{equation}
	Recall = \frac{1}{q} \sum\nolimits_{i=1}^{q} \frac{R_i}{T_i} \notag , \qquad
	Ratio = \frac {\sum_{i=1}^{q} R_i}{\sum_{i=1}^{q} T_i} \notag ,
\end{equation}
where $R_i$ and $T_i$ denote the number of retrieved and true relevant documents of $i$-th query respectively, and $q$ denotes the number of queries.

Since Cosine can conduct retrieval just based on neural representations, we only count the time to retrieve 1000 documents and not include the time to fetch these documents into memory in Lucene for fair comparison. As for the ParSearch, the time cost consists of two stages: retrieve and merge. We use the maximum time of BM25 and Cosine as the cost of retrieval stage. 

Given the limited number of queries for each collection, we randomly divide them into 5 folds and conduct 5-fold cross-validation.

\subsection{Retrieval Performance and Analysis}
This section presents the performance of different index and search schemes on two benchmark TREC collections, and the summary of results is displayed in Table~\ref{tab:text_recall}. Here, we adopt Cosine$_{1000}$ for ParSearch since Cosine$_{1000}$ performs the most precise $k$-NN search with a lower efficiency cost than BM25.

While comparing the performance of the recall@1000 metric, we can see that the neural index based method Cosine performs significantly worse than the symbolic index based method BM25 on two collections. This result suggests that exact matching signal plays an important role in IR, and neural representations could not well capture this signal. Meanwhile, ParSearch and SeqSearch both can achieve better effectiveness consistently, demonstrating that neural representations can be helpful to improve the recall of relevant documents. We find SeqSearch can achieve better performance than ParSearch consistently. The reason is that semantic matching with neural representation alone brings many noise matching signals as we have discussed in Section 3.5. 

Although $k$-NN search in non-metric space is an interesting idea, we find LinComb performs extremely worse on the IR benchmark collections. \citet{boytsov2016off} point out that NAPP is effective only if comparing a query and a document with the same pivot provides meaningful proximity information. In our IR experiments, the queries are extremely short. For example, the title and description of Robust04 are 2.63 and 8.16 on average correspondingly. The short query makes it hard to have overlap terms with limited pivots, leading to less effectiveness of BM25. Meanwhile, the cosine similarity is too coarse and comparing the query and its relevant documents with a nearly random pivot in the semantic space provides confusing information. 

Comparing the results on two collections, we find that the improvement by ParSearch and SeqSearch on WT2G is not so significant as on Robust04. The reason is that BM25 has achieved very good performance, leading to little space for further improvement. We will conduct detailed analysis later in Section 4.5.

When we look at SeqSearch, it is surprising that although only the documents associated by 25\% of the seeds are computed, the effectiveness is not affected too much. This is consistent with our intuition that the documents semantically similar to better exactly matched seeds have higher chances to be relevant. The detailed analysis will be conducted in Section 4.5.

As for the efficiency, we find that Cosine can achieve the best performance, while ParSearch and SeqSearch cost more time than BM25 due to the additional process. Considering the improvement of the effectiveness, the efficiency overhead could be acceptable. We can see the relative additional cost of our search schemes both become smaller as the query length increases. For example, the title and description of Robust04 are 2.63 and 8.16 on average, while the additional cost of ParSearch are 14.22\% and 2.32\% of BM25 correspondingly. The reason is that longer descriptions lead to more documents being computed in symbolic search based on the inverted index, while our additional cost keep almost constant as the query length increases. Meanwhile, Cosine could cost almost the same time, since the query is always the 200-dimension vectors no matter what the original query length is. 

Surprisingly, we find LinComb can not achieve good efficiency even at low recall. The reason is also due to the too short query, which leads to a lot of ineffective computations as many shared $k$-NPs are meaningless. In contrast, \citet{boytsov2016off} retrieve the best answer using the question summary concatenated with the description. They report that $k$-NN search based on NAPP can be more than 1.5x faster than Lucene on Stack Overflow dataset \footnote{The dump from https://archive.org/download/stackexchange dated March 10th 2016} whose query length is 48.4 on average. But on Yahoo!Answers Comprehensive dataset\footnote{https://webscope.sandbox.yahoo.com} whose query length is only 17.8, their approach is 2x slower than Lucene when achieving about the same recall.

\subsection{Analysis on Retrieved Relevant Documents}
In this section, we take the retrieval using topic titles as example, and conduct detailed analysis on retrieved relevant documents to study the utility and difference of both symbolic and neural indices.

\begin{table}[h]
\centering
\caption{Statistics of the averaged relevant documents retrieved by BM25 and Cosine$_{1000}$ using topic titles.}
\label{tab:bm25_cosine_relevant}
\begin{tabular}{ccc}
\hline
 & Robust04 & WT2G \\
\hline
Intersection of BM25 and Cosine$_{1000}$ & 21.81 & 21.42\\ 
Only by BM25 & 18.47 & 16.06 \\
Only by Cosine$_{1000}$ & 8.00 & 2.52\\
\hline
\end{tabular}
\end{table}

Since the results of ParSearch are from both BM25 and Cosine$_{1000}$, we analyze the relevant documents retrieved by BM25 and Cosine$_{1000}$ separately, and the statistics are shown in Table~\ref{tab:bm25_cosine_relevant}. As we can see, more than half of the relevant documents by BM25 and Cosine$_{1000}$ are the same, while others can only be retrieved by BM25 or by Cosine$_{1000}$. For example, on Robust04, 18.47 documents on average found by BM25 are not in the results of Cosine$_{1000}$. Meanwhile, Cosine$_{1000}$ can find 8.00 relevant documents missed by BM25. Note that the relevant documents retrieved only by Cosine$_{1000}$ is not too much on WT2G, thus leading to less improvement on WT2G. We can find that BM25 could find more relevant documents than Cosine$_{1000}$, this suggests the importance of exact matching. 

\begin{table}[h]
\centering
\caption{Statistics of the averaged relevant documents retrieved by seed and associated documents using topic titles.}
\label{tab:association_25_recall}
\begin{tabular}{ccc}
\hline
  & Robust04 & WT2G \\
\hline
Seed Documents & 680 & 800 \\
Relevant Seed Documents & 36.91 & 36.44 \\
\hline
Associated Documents with 25\% Seeds & 1773 & 1577 \\ 
Relevant Associated Documents & 10.28 & 3.76\\
\hline
Associated Documents with 50\% Seeds & 3441 & 2977 \\ 
Relevant Associated Documents & 12.80 & 4.60 \\
\hline
Associated Documents with 100\% Seeds &  6666 & 5534 \\ 
Relevant Associated Documents &  15.80 & 5.58 \\
\hline
\end{tabular}
\end{table}

As for the SeqSearch, the number of relevant documents contained in seed and associated documents are shown in Table~\ref{tab:association_25_recall}. As we can see, the $k$-NNs in the semantic space of well exactly matched documents contains many relevant documents, e.g.,  10.28 relevant documents in the 20-NNs of $700*25\% = 175$ seeds on Robust04. Note that we remove the seed documents from the initially associated documents, so the number of relevant documents may be larger in the 20-NNs of seeds. We can see that on WT2G, the proportion of relevant documents found in associated documents is not as big as on Robust04. Since exact matching has worked well enough on WT2G, the improvement on WT2G is thus not significant. Moreover, using a high proportion of seeds to associate can find a little more relevant documents while the number of associated documents increases exponentially, leading to more difficulty to distinguish the relevant documents with cosine similarity. This is the reason why SeqSearch$_{25\%}$ can achieve comparable performance with SeqSearch$_{100\%}$.

\section{Conclusions}
In this paper, we argue that neural representations can also be employed to improve the recall of relevant documents for initial retrieval. To solve the index and search challenges, we introduce a $k$-NN graph based neural index and further propose the parallel search scheme and the sequential search scheme based on both neural and symbolic indices. Our experiments show that both hybrid index and search schemes can improve the recall of the initial retrieval stage with small overhead.

For future work, we would like to conduct empirical study of complex similarity in $k$-NN graph which may achieve better effectiveness. Since Cosine can achieve the best efficiency, we also would like to explore the other sequential way, i.e., retrieve by Cosine in the semantic space firstly and conduct association process in the symbolic space.